\documentclass[12pt]{article}
\usepackage{amsmath}
\usepackage[dvips]{graphicx}
\newfont{\sfb}{cmssbx10 scaled 1400}
\newfont{\bigsf}{cmssbx10 scaled 1600}
\renewcommand{\thefootnote}{\fnsymbol{footnote}}
\allowdisplaybreaks[1]

\begin{document}
\begin{flushright}
FUT--07--01\\
May~~~~2007
\end{flushright}
\begin{center}
{\baselineskip 26pt
{\LARGE\bf $F$ and $D$ Values with Explicit Flavor Symmetry Breaking and $\Delta s$
 Contents of Nucleons}\\

\vspace{2.0em}

{
\renewcommand{\thefootnote}{\fnsymbol{footnote}}
T. Yamanishi\\

\vspace{0.8em}

{\it Department of Management Information Science,}\\
{\it Fukui Institute of Technology, Gakuen, Fukui 910--8505, Japan}\\

\vspace{3.5em}
}
}

{\bf Abstract}
\end{center}

\baselineskip=25pt

We propose a new model for describing baryon semi-leptonic decays for estimating $F$ and $D$
 values with explicit breaking effects of both SU(3) and SU(2) flavor symmetry,
 where all possible SU(3) and SU(2) breaking effects are induced from an 
 effective interaction.
An overall fit including the weak magnetism form factor yields $F=0.477\pm 0.001$,
 $D=0.835\pm 0.001$, $V_{ud}=0.975\pm 0.002$, and $V_{us}=0.221\pm 0.002$ with
 $\chi^2=4.43/5$ d.o.f.
The spin content of strange quarks $\Delta s$ is estimated from the obtained values
 $F$ and $D$, and the nucleon spin problem is re-examined.
Furthermore, the unmeasured values of $(g_1/f_1)$ and $(g_1)$ for other hyperon
 semi-leptonic decays are predicted from this new formula.
\vspace{1.0em}

\noindent

PACS number(s): 13.30.Ce, 11.30.Hv, 14.20.Dh

\vfill\eject

\noindent

\section{Introduction}

Triggered by the measurement of the polarized structure function of proton
 $g^p_1(x)$ by EMC in 1987\cite{EMC}, the internal structure of a nucleon remains a
 challenging subject in nuclear and particle physics.
Surprisingly, the measured $g^p_1(x)$ implies that only
 a small part of the nucleon spin is carried by quarks,
 and the polarization of the strange quark is negative and quite
 large.
At the leading-order of QCD, the amount of the strange quark carrying
 the nucleon spin $\Delta s(Q^2) (\equiv\Delta s_s+\Delta\bar s_s)$ is given by
 \begin{equation}
   \Delta s(Q^2)= 3~\Gamma_1^p(Q^2)-\frac{1}{4}g_A-
   \frac{5}{12}(3F-D)~,
   \label{eqn:Delta_s}
 \end{equation}
 where $F$ and $D$ are Cabibbo parameters\cite{Cabi63} and $\Gamma_1^p(Q^2)$ is
 the first moment of the proton structure function $g^p_1(x)$.
$g_A$ is the ratio of axial-vector to vector coupling constants
 ( equivalent to the axial-vector to vector form factor $g_1/f_1$ )
 of the neutron $\beta$ decay.
The values of $F$ and $D$ can be uniquely determined by the ratio of
 the axial-vector to vector form factor $g_1/f_1$ of various baryon semi-leptonic
 decays.
The EMC value of $g^p_1(x)$ leads to
 \begin{equation}
   \Delta s(Q^2=10.7{\rm GeV}^2)=-0.190\pm 0.032\pm 0.046~,
   \label{eqn:Delta_s_EMCvalue}
 \end{equation}
using $F$ and $D$ values with the assumption of SU(3) flavor symmetry.
This result is not anticipated in conventional quark models,
 and is often referred to as {\it the proton spin crisis} \cite{Lampe}.
However, note that the SU(3) flavor symmetry is not
 a good description because of the rather large mass difference
 between strange and nonstrange quarks.
Furthermore, recent data on the longitudinally polarized semi-inclusive
 DIS \cite{HERMES} suggest the asymmetry between $\Delta\bar u(x)$
 and $\Delta\bar d(x)$; that is , even SU(2) flavor is broken.

Now, significant deviations between high precision data on various
 baryon semi-leptonic decays and $g_1/f_1$ expressed from
 $F$ and $D$ under SU(3) flavor symmetry are seen.
Accordingly, it is important to reconsider the result of
 eq.(\ref{eqn:Delta_s_EMCvalue}) using $F$ and $D$ values without
 SU(3) and SU(2) flavor symmetry.

There have been several attempts to estimate $F$ and $D$ values
 by taking the SU(3) or SU(2) flavor breaking into account\cite{Ratcliffe04}.
For the SU(3) flavor, the breaking effects have been considered from
 the effective Hamiltonian formalism \cite{Ratcliffe97,Ratcliffe99},
 possible SU(3) breaking with hypercharge matrix
 $\lambda_8$ \cite{Song},
 the 1/N$_c$ expansion \cite{Dai,Mendieta}, the baryon mass differences
 \cite{Ehrnsperger}, and the center-of-mass corrections \cite{Donoghue}.  
In the case of SU(2), $F$ and $D$ values have been estimated by considering the
 $\Lambda ^0$-$\Sigma ^0$ mixing induced from the isospin-violation \cite{Karl}.
Although these analyses are interesting in themselves, they are not yet
 general and are insufficient to constrain the parameter of polarized quark
 distribution functions describing quark spin contents. 
In this paper, to estimate more reliable $F$ and $D$ values,
 we propose a new model of baryon semi-leptonic decays that incorporates
 both SU(3) and SU(2) flavor symmetry breaking generally,
 and we attempt to derive the polarized strange-quark content in the proton
 from those $F$ and $D$ values.

In the next section, we derive the most general formulas for $F$ and $D$ values
 without SU(3) and SU(2) flavor symmetry.
The $\chi^2$ analysis is carried out in Section 3.
Using the obtained $F$ and $D$ values, we estimate the first moment
 of polarized strange-quarks.
Section 4 provides the summary and discussion.

\section{Form factors with both SU(3) and SU(2) flavor symmetry breaking}

The matrix element for baryon semi-leptonic decay $A\to B+\ell + \bar\nu$
 is given by
 \begin{equation}
   \mathcal{M}=\frac{G_F}{\sqrt{2}}<B|J_h^{\mu}|A>
   \bar{u_{\ell}}(p_{\ell})\gamma_{\mu}(1+\gamma_5)u_{\nu}(p_{\nu})~,
   \label{eqn:m-e}
 \end{equation}
 where $G_F$ is the universal weak coupling constant, and
 $p_{\ell}$ and $p_{\nu}$ the four momenta of the lepton and anti-neutrino,
 respectively.
The hadronic current is written as\cite{Renton}
 \begin{eqnarray}
   <B|J_h^{\mu}|A>&=&C\bar{u_B}(p_B)\left\{f_1(q^2)\gamma^{\mu}+
   i\frac{f_2(q^2)}{M}\sigma^{\mu\nu}q_{\nu}+\frac{f_3(q^2)}{M}q^{\mu}+ \right .
   \label{eqn:b-m-e}\\
   & &g_1(q^2)\gamma^{\mu}\gamma_5+i\frac{g_2(q^2)}{M}\sigma^{\mu\nu}q_{\nu}\gamma_5
   +\left . \frac{g_3(q^2)}{M}q^{\mu}\gamma_5\right\}u_A(p_A)~, \nonumber
 \end{eqnarray}
where $C$ is the mixing parameter $V_{ud}$ or $V_{us}$ with the
 Cabibbo-Kobayashi-Maskawa (CKM)
 matrix element for $|\Delta S|=0$ or $1$ transitions, respectively.
$p_A$ and $p_B$ are the four momenta of the initial and final baryons, and $M$ the mass of
 the initial baryon. 
$q$ is the momentum transfer given by $q=p_A-p_B$.
The functions $f_i(q^2)$ and $g_i(q^2)$ with $i=1$, $2$, and $3$ are the vector
 and axial-vector current form factors, respectively, and include all information on
 hadron dynamics.  
In literature, $f_1$, $f_2$, and $f_3$ are called the vector, the induced tensor or
 weak magnetism, and the induced scalar form factors, respectively, and
 $g_1$, $g_2$, and $g_3$ are the axial-vector, the induced pseudo-tensor or
 weak-electricity, and the induced pseudoscalar form factors, respectively.
In Weinberg's classification, $f_3$ and $g_2$ are associated with second-class
 currents, whereas the others are first-class currents\cite{Weinberg}.

Now, we consider the Hamiltonian 
 \begin{equation}
   {\mathcal H}={\mathcal H}_0+h'~,
 \end{equation}
 describing mass-splitting interactions, where ${\mathcal H}_0$ is SU(3) symmetric, and
 $h'$ is responsible for the mass differences among particles in baryon or meson
 SU(3) multiplets\cite{Brene,Sirlin}.
Assuming that ${\mathcal H}$ is invariant under charge conjugation, and that
 the breaking term $h'$ is originated from the third and
 eighth components of an octet, we can expand
 the weak current to the first order in $h'$, as \cite{Ademollo}
 \begin{eqnarray}
   &&a_0~{\rm Tr}(\bar BB\lambda_i)+b_0~{\rm Tr}(\bar B\lambda_iB)+
   a~{\rm Tr}(\bar BB\{\lambda_i, (\alpha\lambda_3+\beta\lambda_8)\})+
   b~{\rm Tr}(\bar B\{\lambda_i, (\alpha\lambda_3+\beta\lambda_8)\}B)
   \nonumber\\
   &&+c~[{\rm Tr}(\bar B\lambda_iB(\alpha\lambda_3+\beta\lambda_8))-
   {\rm Tr}(\bar B(\alpha\lambda_3+\beta\lambda_8) B\lambda_i)]
   +g~{\rm Tr}(\bar BB){\rm Tr}(\lambda_i(\alpha\lambda_3+\beta\lambda_8)) \label{eqn:AG} \\
   &&+h~[{\rm Tr}(\bar B\lambda_i){\rm Tr}(B(\alpha\lambda_3+\beta\lambda_8))+
   {\rm Tr}(\bar B(\alpha\lambda_3+\beta\lambda_8)){\rm Tr}(B\lambda_i)]~,
   \nonumber
 \end{eqnarray}
 for the first-class covariants of $\gamma_{\mu}$, $\sigma_{\mu\nu}q_{\nu}$,
 $\gamma_{\mu}\gamma_5$, and $\gamma_5q_{\mu}$.
Here, $i$ gives the $i$-th component of the weak current, and 
$\alpha$ and $\beta$ are taken as parameters for SU(2) and SU(3) symmetry breaking
 effects, respectively.
$a_0$, $b_0$, $\cdots$, $h$ are the first-class amplitudes.
In particular, at zero, four-momentum transfer $q^2\to 0$, eq.(\ref{eqn:AG})
 is reduced to its first two terms, where $a_0=D-F$ and $b_0=D+F$ with
 $F$ and $D$ in the SU(3) symmetry limit.
Note that the SU(2) symmetry is realized only when $\alpha=0$.
$B$ in eq.(\ref{eqn:AG}) is the matrix representing the baryon octet:
 \begin{equation}
   B=\left[
     \begin{array}{ccc}
      \frac{1}{\sqrt{2}}\Sigma^0+\frac{1}{\sqrt{6}}\Lambda^0 & \Sigma^+ & p\\ 
      \Sigma^- & -\frac{1}{\sqrt{2}}\Sigma^0+\frac{1}{\sqrt{6}}\Lambda^0 & n \\
      \Xi^- & \Xi^0 & -\frac{2}{\sqrt{6}}\Lambda^0
     \end{array}
   \right]~.
   \label{eqn:B-octet}
 \end{equation}

\vspace{1em}

\subsection{The vector and axial-vector form factors $f_1$ and $g_1$}

The Ademollo-Gatto theorem guarantees that the vector form factor $f_1$
 is not modified in the first order in the symmetry breaking, but occurs in
 the second order\cite{Ademollo}.
In contrast, the axial-vector form factor $g_1$ is affected by 
 the first-order symmetry breaking. 
In the present analysis, we only consider the corrections from the first-order
 symmetry breaking of SU(2) and SU(3).
After some algebraic calculation with eq.(\ref{eqn:AG}),
 we obtain the ratios of the axial-vector to vector
 form factors $g_1/f_1$ as
\begin{subequations}
 \begin{align}
& (g_1/f_1)_{n\to p}=F+D+2\beta(b-c)~, \label{eqn:np}\\
& (g_1/f_1)_{\Lambda^0\to p}=F+D/3-(\alpha-\beta)(a-2~b)/3+(\alpha-3\beta)~c/3-
  2\beta h/3~, \label{eqn:Lp}\\
& (g_1/f_1)_{\Sigma^-\to n}=F-D-(\alpha-\beta)(a+c)~, \label{eqn:Sn}\\
& (g_1/f_1)_{\Xi^-\to\Lambda^0}=F-D/3-(\alpha-\beta)(2~a-b)/3+(\alpha-3\beta)~c/3+
  2\beta h/3~, \label{eqn:XL}\\
& (g_1/f_1)_{\Xi^0\to\Sigma^+}=F+D+(\alpha-\beta)(b-c)~, \label{eqn:XS}\\
& (g_1)_{\Sigma^-\to\Lambda^0}=\sqrt{\frac{2}{3}}\left\{ D+\beta(a+b)+
  \alpha c+3\beta h\right\}~, \label{eqn:SL}\\
& (g_1/f_1)_{\Xi^-\to\Xi^0}=F-D-2\beta(a+c)~, \label{eqn:XX}\\
& (g_1/f_1)_{\Xi^-\to\Sigma^0}=F+D+(\alpha-\beta)b+(\alpha+\beta)c+2\alpha h~, \label{eqn:XMS}\\
& (g_1)_{\Sigma^+\to\Lambda^0}=\sqrt{\frac{2}{3}}\left\{ D+\beta(a+b)-
  \alpha c+3\beta h\right\}~, \label{eqn:SPL}
 \end{align}
\end{subequations}
 with the first-class amplitudes $a$, $b$, $c$, $h$ of the correction terms
 and $\alpha$, $\beta$ describing
 the contributions of SU(2) and SU(3) symmetry breaking.
Since the vector form factor is absent in $\Sigma^-\to\Lambda^0$ and $\Sigma^+\to\Lambda^0$,
 we present the expression for the axial-vector form factor.
In the absence of the breaking parameters $\alpha$ and $\beta$ for SU(2) and SU(3) symmetries,
 $i.e.$, the SU(3) limit,
 we see that eqs.(\ref{eqn:np})--(\ref{eqn:SPL}) are reduced to the formulas given by
 the Cabibbo model.

\vspace{1em}

\subsection{The weak magnetism form factor $f_2$ and others including $f_3$, $g_2$, and $g_3$}

The induced tensor or weak magnetism form factor $f_2$ at $q^2=0$
 for each baryon semi-leptonic decay is obtained in terms of the proton and
neutron anomalous magnetic moments in the Cabibbo model with SU(3) limit\cite{Renton}.
When the SU(3)/SU(2) symmetry is broken, $f_2$ would be corrected by 
 the symmetry breaking effects.
Although the electron spectrum in baryon semi-leptonic decays is
 sensitive to the magnitude of $f_2/f_1$ \cite{Garcia85,Cabi03},
 the measurement of $f_2$ has not yet been carried out experimentally with
 sufficient precision.
Traditionally, the symmetry effect is expressed by multiplying the
 $M/M_p$ factor from the SU(3) symmetry parameters represented by
 the Cabibbo model at $q^2=0$, where
 $M$ and $M_p$ are the mass of the decaying baryon $A$ and the
 proton, respectively.
However, we find a significant deviation in the model prediction for symmetry
 breaking $f_2/f_1=\frac{1}{2}\frac{M_{\Sigma^-}}{M_p}(\mu_p+2\mu_n)=-1.297$ from
 the experimental data $-0.96\pm0.15$ for $\Sigma^-\to n\ell\bar\nu_{\ell}$ \cite{Hsueh}.
Here, we express $f_2$ for various decay processes in the symmetry
 breaking case in terms of the anomalous magnetic moments of relevant baryons 
 instead of the proton $\mu_p$ and neutron $\mu_n$ anomalous magnetic moments.
The expressions of $f_2$ are given in Table 1 for three cases: SU(3)
 symmetry, SU(3) symmetry breaking, and both SU(3) and SU(2) symmetry
 breaking.
The $f_2/f_1$ for $\Sigma^-\to n\ell\bar\nu_{\ell}$ in the symmetry breaking case
 becomes $-0.8765$, and is consistent within the range of error with the
 experimental data.

The weak-electricity form factor $g_2$ vanishes
 at the SU(3) limit and V-spin invariance in the absence of second-class
 current.
The induced scalar form factor $f_3$ also goes to zero in SU(3) limit.
In the real world the SU(3) symmetry is broken and therefore
 these form factors are expected to have some significant value.
However, we ignore the $g_2$ terms because
 the corrections to $g_2$ in first-order symmetry breaking 
 contribute to the decay amplitude in the second order\cite{Garcia92,Mateu}.
Furthermore, since the contribution of $f_3$ to baryon semi-leptonic decay is suppressed by the
 mass of the emitted lepton, we can safely omit the term including $f_3$.
Since the induced pseudoscalar form factor $g_3$ is also proportional to
 $m_{\ell}$, this contribution is negligible as well.

\section{Numerical analyses and results}

We determine the parameters of $g_1/f_1$ given in eqs.(\ref{eqn:np})--(\ref{eqn:SPL})
 and the CKM matrix elements $V_{ud}$ and $V_{us}$ from $\chi^2$-analysis
 with experimental data for ratios of the axial-vector to vector form factors
 and for differential semi-leptonic decay rates of baryons.
Then, we obtain the values of the parameters for each case of symmetry
 and symmetry breaking $-$SU(3) symmetry, SU(3) symmetry
 breaking, and both SU(3) and SU(2) symmetry breaking. 
Table 2 shows the data used here\cite{PDG}.
Note that the KTeV group implied two values for $g_1/f_1$ of
 $\Xi^0\to\Sigma^+\ell\bar\nu$ \cite{KTeV}.
One is the case of SU(3) symmetry, and its value is
  $1.32\pm ^{0.22}_{0.18}$.
The other is the SU(3) breaking case: $1.17\pm 0.28\pm 0.05$.
We adopt the former for analysis of SU(3) symmetry case and the latter for
 symmetry breaking cases.

In fitting our expression for the decay rate to experimental data,
 we include not only the form factors $f_1$ and $g_1$
 but also $f_2$.
We also take into account the radiative corrections and the $q^2$-dependence
 of these form factors.

Radiative corrections play an important role in the precise determination of
 the CKM matrix, and hence should not be
 neglected in estimating the effects of SU(3) symmetry breaking\cite{Bourquin83}.
There are various sets of radiative corrections to the semi-leptonic
 decay emitting an electron, but the differences in the differential cross section
 obtained using them varied only by a few percent\cite{Bourquin83}.
Here, we adopt a set of radiative corrections for the hyperon semi-leptonic
 decays calculated by T{\' o}th et al.\cite{Toth} shown in Table 3. 
The radiative corrections for $\Xi^-\to\Sigma^0e\bar\nu$ and
 $\Xi^0\to\Sigma^+e\bar\nu$, which are not estimated by ref.\cite{Toth},
 are simply assumed to be the same as that for $\Xi^-\to\Lambda e\bar\nu$

For the $q^2$-dependence we use the dipole form 
  \begin{eqnarray}
   f_i(q^2)&=&\frac{f_i(0)}{(1-q^2/M_V^2)^2}~,
    \label{eqn:q2-dep} \\
   g_i(q^2)&=&\frac{g_i(0)}{(1-q^2/M_A^2)^2}~,
   \nonumber
 \end{eqnarray}
 for $i=1, 2$ with $M_V=0.84~(0.97)$GeV and $M_A=1.08~(1.25)$GeV at
 $|\Delta S|=0~(1)$ decays, respectively\cite{Cabi03}.
As the physical range of $q^2$ is given in $m^2_{\ell}\leq q^2 \leq (M-M')^2$,
 and its value is smaller than $M_{V, A}^2$,
 we approximate the products of the form factors as follows:
  \begin{equation}
   f_i(q^2)f_j(q^2)=f_i(0)f_j(0)(1+4q^2/M_{V, A}^2)~.
    \label{eqn:q2-appx}
  \end{equation}
The contributions of the term $4q^2/M_{V, A}^2$ in eq.(\ref{eqn:q2-appx})
 to the differential cross section
 for $f_2^2$ and $f_1f_2$ are an order of $10^{-2} - 10^{-3}$ less
 than those for $f_1^2$ and $g_1^2$.
Then, we ignore the term $4q^2/M_{V, A}^2$ of $f_2^2$ and $f_1f_2$ in our
 analysis.

The $\chi^2$ values for fitting are shown in Table 4, where
 we see that the $\chi^2$ values of SU(3) symmetry
 for the decay rates of
 $\Xi^-\to\Lambda^0e\bar\nu$ and $\Xi^-\to\Sigma^0e\bar\nu$
 are significantly improved by considering the symmetry breaking effects.
Furthermore, we could say that not only SU(3) but also SU(2) symmetry breaking effects
 should be considered to reproduce the experimental data beyond the Cabibbo model with
 the SU(3) symmetry since the $\chi^2/$d.o.f. value for both SU(3) and SU(2) breaking
 is smaller than for SU(3) breaking alone.

Table 5 lists the optimal parameters for each case.
In the present analysis, the parameters $a$, $b$, $c$ and $h$ are smaller than
 $F$ and $D$; thus
 a perturbative expansion with our effective interaction is reliable.
From these results, we obtain $F/D=0.599\pm 0.006$ for the SU(3) symmetry,
 $0.648\pm 0.004$ for the SU(3) breaking,
 and $0.572\pm 0.010$ for both SU(3) and SU(2) breaking, respectively.
The effects of both SU(3) and SU(2) symmetry breaking make
 $F/D$ rather small.
Figure 1 shows the $F$ and $D$ values calculated for
 SU(3) symmetry and for both SU(3) and SU(2) breaking,
 where the bands of allowed $F$ and $D$ values are deduced from 
 $(g_1/f_1)$ measurements.
For the SU(3) symmetry case (Fig.1(a)), the bands that
 present the experimental errors do not cross at a point.
However, Fig.1(b) shows that the bands completely corrected by the effects of
 both SU(3) and SU(2) symmetry breaking share a common overlapping region,
 and one can determine the $F$ and $D$ values in this overlapping region.

These results for $\chi^2$ values (Fig.1) imply the contribution of SU(2) flavor
 symmetry breaking on baryon semi-leptonic decays.
However, should the SU(2) breaking effect for these processes actually be
 considered ?
We consider the
$\Sigma^-\to\Lambda^0\ell\bar\nu$ process because in that process only
 $|\Delta S|=0$ interactions work, and one can easily test the presence or absence of
 SU(2) symmetry breaking effects for baryon semi-leptonic decays.
Note that our eqs.(\ref{eqn:SL}) and (\ref{eqn:SPL}), only in the case of SU(3) symmetry
 breaking but SU(2) symmetry is kept, lead to a natural result that 
 $(g_1)_{\Sigma^-\to\Lambda^0}$ and $(g_1)_{\Sigma^+\to\Lambda^0}$ degenerate.
Our prediction for the axial-vector form factor $(g_1)$ at
 $\Sigma^-\to\Lambda^0\ell\bar\nu$ becomes
\begin{align}
  g_1~\cos\theta_c &=0.632\pm 0.004~~~~{\rm for~~the~~SU(3)~~symmetry~~case}, \nonumber\\ 
                   &=0.593\pm 0.038~~~~{\rm for~~the~~SU(3)~~breaking~~case}, \nonumber\\
                   &=0.586\pm 0.040~~~~{\rm for~~the~~SU(3)~~and~~SU(2)~~breaking~~case}, \nonumber
\end{align}
 with the Cabibbo angle $\cos\theta_c$ using each value of $V_{ud}$ ($\equiv \cos\theta_c$). 
This $(g_1)_{\Sigma^-\to\Lambda^0}$ was already
 measured in the WA2 experiment \cite{Bourquin82}\footnote[3]{The experimental result
 is not used to determine our parameters by $\chi^2$ fitting, because it is not listed in the
 present particle data table.}~:
\begin{equation}
  g_1~\cos\theta_c = 0.572\pm 0.016~.
  \label{eqn:BGHORS}
\end{equation}
The case of both SU(3) and SU(2) breaking is in agreement with the
 data of eq.(\ref{eqn:BGHORS}) within the error boundary.
Experimental data indicate the importance of corrections due to SU(2)
 breaking.

Using our results, one can predict the unmeasured ratios of axial-vector to vector
 form factors 
 $(g_1/f_1)_{\Xi^-\to\Xi^0}$, $(g_1/f_1)_{\Xi^-\to\Sigma^0}$
 and the axial-vector form factor 
 $(g_1)_{\Sigma^+\to\Lambda^0}$
 of the hyperon semi-leptonic decay.
The $\Sigma^+\to\Lambda^0\ell^+\nu$
 process is interesting as well, because in that process only
 $|\Delta S|=0$ interactions work the same as when $\Sigma^-\to\Lambda^0\ell\bar\nu$,
and we can easily test whether the presence or absence of
 SU(2) symmetry breaking is effective.
In the case of both SU(3) and SU(2) flavor symmetry breaking, we get
\begin{align}
  (g_1/f_1)_{\Xi^-\to\Xi^0} &=-0.144\pm 0.082~, &
  (g_1/f_1)_{\Xi^-\to\Sigma^0} &=1.283\pm 0.033~, \nonumber\\
  (g_1)_{\Sigma^+\to\Lambda^0}&=0.668\pm 0.041~, & & \nonumber
\end{align}
 by using eqs.(\ref{eqn:XX})--(\ref{eqn:SPL}), respectively.
The measurement of unmeasured $(g_1/f_1)$ and $(g_1)$ is important to
 determine the magnitude of the symmetry breaking effect and test
 the validity of our model, which hopefully will be carried out in the near future.

Now, we attempt to estimate the amount of strange quark content
 carrying the proton spin using our $F$ and $D$ values. 

From the recent HERMES result $\Gamma_1^d(Q^2=5{\rm GeV}^2)=0.0437\pm0.0035$\cite{HERMES06}
 and our $F$ and $D$ values, we have
 \begin{eqnarray}
   \Delta s &=& \Delta s_s+\Delta\bar s_s
             = \frac{3\Gamma_1^d}{1-\frac{3}{2}\omega_D}-\frac{5}{12}(3F-D)\nonumber\\
            &=& -0.107\pm 0.015~,
   \label{eqn:Delta_s_value_gd}
 \end{eqnarray}
 for both SU(3) and SU(2) symmetry breaking with the D-state admixture to the
 deuteron wave function, $\omega_D=0.05\pm 0.01$
 at the leading-order of QCD, while
 \begin{equation}
   \Delta s=-0.122\pm 0.013~,
   \label{eqn:Delta_s_value_mod}
 \end{equation}
 for SU(3) symmetry.
Thus, the central value of $\Delta s$ for symmetry breaking
 increases by about $12$\% compared with the symmetry case.
Therefore, the flavor breaking effect contributes significantly to quark spin
 contents.
This suggests that re-analysis of the proton spin structure using $F$ and $D$
 values with full breaking effects is necessary for solving the proton spin crisis.

\section{Summary and discussion}

We have proposed a new model of baryon semi-leptonic decays to derive 
 $F$ and $D$ with both SU(3) and SU(2) flavor symmetry breaking effects.
Numerical analysis using present experimental data determined that
 $F$ and $D$ values with both SU(3) and SU(2) symmetry
 breaking effects described the experimental data well, rather than
 those with the exact SU(3) symmetry case.
The $\chi^2$ fit leads to values of $F/D$ in the case of both
 SU(3) and SU(2) breaking that becomes smaller by about $5$\% than that of the SU(3)
 symmetry case.

The central value for the amount of the strange quarks carrying the nucleon spin
 $\Delta s$ for both SU(3) and SU(2) breaking cases increases by about
 $12$\% compared with the one for the SU(3) symmetry case.
Therefore, it is very important to re-analyze the 
 polarized parton distribution functions using $F$ and $D$ values with SU(3)/SU(2)
 symmetry breaking effect.
The Japan Proton Accelerator Research Complex (J-PARC) is now under construction.
It provides high intensity neutrino beams via charged pion decays from 
 50 GeV high intensity proton and antiproton beams.
One expects that direct measurements of $\Delta s$ for the nucleon scattering off
 the neutrino will be conducted at J-PARC with high precisions.

After completion of this work, flavor SU(3) breaking effects in quenched
 lattice QCD simulations by Sasaki et al.\cite{Sasaki} were brought to my attention.
They reached a similar result for the $V_{us}$ value.


\vfill\eject

\begin{center}
\Large\bf{Acknowledgments}
\end{center}
I thank T. Morii and K. Ohkuma for enlightening comments
 and for careful reading of the manuscript.
This work was supported by FUT Research Promotion Fund.

\begin{figure}
 \begin{center}
  \includegraphics[width=14.5cm,clip]{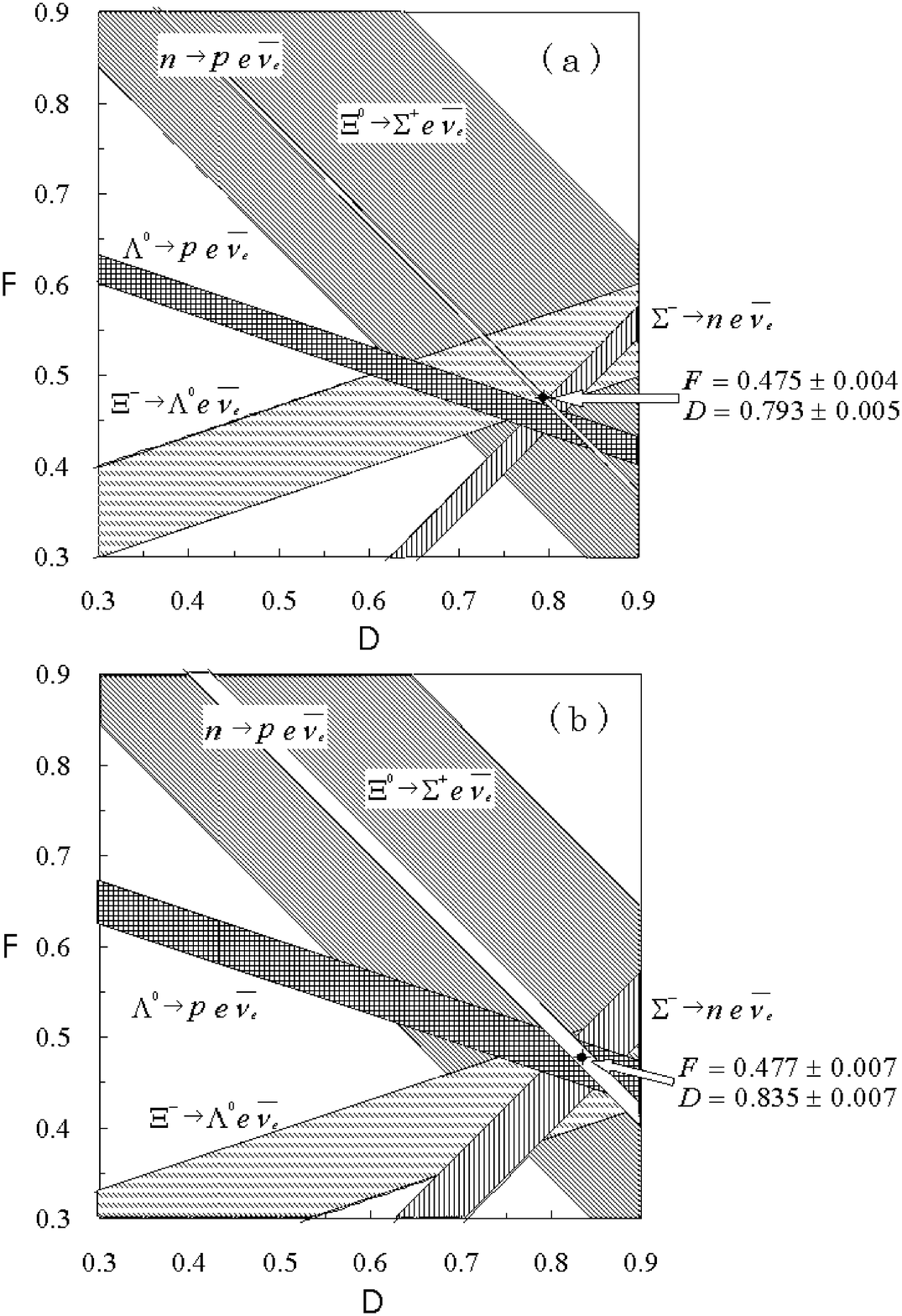}
 \end{center}
\caption{The allowed area of $F$ and $D$ obtained from $\chi^2$ fit to the measurements
 of $(g_1/f_1)$ for various baryon semi-leptonic decays is shown for two cases:
 (a) SU(3) symmetry and (b) both SU(3) and SU(2) breaking.
The black circle in each figure indicates the optimal value of $F$ and $D$ for
 the fit.
Each band indicates the experimental errors for SU(3) symmetry in (a).
 On the other hand, it is slightly modified by the breaking parameters in
 eqs.(\ref{eqn:np})--(\ref{eqn:SPL}) for both SU(3) and SU(2) breaking in (b).}
\label{fig:fd}
\end{figure}

\vfill\eject
\newpage

\vspace{3em}

\begin{table}
 \begin{center}
  \begin{tabular}{cccc}
   \hline
       & SU(3) symmetry case  & SU(3) breaking case &
           SU(3) and SU(2) \\
       & & & breaking case \\ \hline
      $n\to p\ell\bar\nu$ & 
      $\frac{1}{2}(\mu_p-\mu_n)=1.853$ &
      $\frac{1}{2}\frac{M_N}{M_N}(\mu_p-\mu_n)=1.853$ &
      $\frac{1}{2}\frac{M_n}{M_p}(\mu_p-\mu_n)=1.855$ \\
      $\Lambda^0\to p\ell\bar\nu$ &
      $\frac{1}{2}\mu_p=0.896$ &
      $\frac{1}{2}\frac{M_{\Lambda}}{M_N}\mu_p=1.065$ &
      $\frac{1}{2}\frac{M_{\Lambda}}{M_p}\mu_p=1.066$ \\
      $\Sigma^-\to n \ell \bar\nu$ &
      $\frac{1}{2}(\mu_p+2\mu_n)=-1.017$ &
      $\frac{1}{2}(\mu_n-\mu_{\Sigma^-})=-0.877$ &
      $\frac{1}{2}(\mu_n-\mu_{\Sigma^-})=-0.877$ \\
      $\Xi^-\to\Lambda^0 \ell \bar\nu$ &
      $-\frac{1}{2}(\mu_p+\mu_n)=0.060$ &
      $-\frac{1}{2}\mu_{\Xi^-}=0.175$ &
      $-\frac{1}{2}\mu_{\Xi^-}=0.175$ \\
      $\Xi^0\to\Sigma^+ \ell \bar\nu$ &
      $\frac{1}{2}(\mu_p-\mu_n)=1.853$ &
      $\frac{1}{2}(\mu_{\Sigma^+}-\mu_{\Xi^0})=1.354$ &
      $\frac{1}{2}(\mu_{\Sigma^+}-\mu_{\Xi^0})=1.354$ \\
      $\Xi^-\to\Sigma^0 \ell \bar\nu$ &
      $\frac{1}{2}(\mu_p-\mu_n)=1.853$ &
      $\frac{1}{2}(4\mu_{\Sigma^0}-\mu_{\Xi^-})=1.123$ &
      $\frac{1}{2}(4\mu_{\Sigma^0}-\mu_{\Xi^-})=1.123$ \\
      $\Sigma^-\to\Lambda^0 \ell \bar\nu$ &
      $-\sqrt{\frac{3}{2}}\frac{1}{2}\mu_n=1.172$ &
      $-\frac{\sqrt{6}}{2}\frac{M_{\Sigma}}{M_{\Lambda}}\mu_{\Lambda}=0.803$ &
      $-\frac{\sqrt{6}}{2}\frac{M_{\Sigma^-}}{M_{\Lambda}}\mu_{\Lambda}=0.806$\\
      $\Sigma^+\to\Lambda^0\bar\ell\nu$ &
      $-\sqrt{\frac{3}{2}}\frac{1}{2}\mu_n=1.172$ &
      $-\frac{\sqrt{6}}{2}\frac{M_{\Sigma}}{M_{\Lambda}}\mu_{\Lambda}=0.803$ &
      $-\frac{\sqrt{6}}{2}\frac{M_{\Sigma^+}}{M_{\Lambda}}\mu_{\Lambda}=0.800$ \\ \hline
  \end{tabular}
 \end{center}
 \caption[Table 1]{$f_2/f_1$ values for three cases, SU(3) symmetry,
 SU(3) breaking, and both SU(3) and SU(2) breaking, are listed.
 For $\Sigma^-\to\Lambda^0 \ell \bar\nu$ and $\Sigma^+\to\Lambda^0\bar\ell\nu$
 processes, the values are given for $f_2$ since $f_1=0$, even though the first-order
 symmetry breaking is taken into account.
 The $\mu_{\Sigma^0}$ of SU(3) breaking and both
 SU(3) and SU(2) breaking for $\Xi^-\to\Sigma^0 \ell \bar\nu$ 
  is taken as $\mu_{\Sigma^0}=(\mu_{\Sigma^+}+\mu_{\Sigma^-})/2$.} 
\end{table}

\begin{table}
 \begin{center}
  \begin{tabular}{cccc}
   \hline
                    & \multicolumn{2}{c}{Rate ($10^6\,$s$^{-1}$)} &  \\ \cline{2-3}        
      \multicolumn{1}{c}{\raisebox{1.5ex}[0pt][0pt]{decay}} &
           $\ell =e$ & $\ell =\mu$ &
          \multicolumn{1}{c}{\raisebox{1.5ex}[0pt][0pt]{$g_1/f_1$}} \\ \hline
      $n\to p\ell\bar\nu$ & $1.129\pm 0.001^*$
          & & $1.2695\pm 0.0029$ \\
      $\Lambda^0\to p\ell\bar\nu$ & $3.16\pm 0.06$ & $0.597\pm 0.133$ & $0.718\pm 0.015$ \\
      $\Sigma^-\to n\ell\bar\nu$ & $6.876\pm 0.236$ & $3.0\pm 0.2$ & $-0.340\pm 0.017$ \\
      $\Xi^-\to\Lambda^0\ell\bar\nu$ & $3.35\pm 0.37$ & $2.1\pm ^{2.1}_{1.3}$ & $0.25\pm 0.05$ \\
      $\Xi^0\to\Sigma^+\ell\bar\nu$ & $0.93\pm 0.14$ & & $1.32\pm ^{0.22}_{0.18}$ or
           $1.17\pm 0.28\pm 0.05$ \\
      $\Xi^-\to\Sigma^0\ell\bar\nu$ & $0.53\pm 0.10$ & & \\
      $\Sigma^-\to\Lambda^0\ell\bar\nu$ & $0.387\pm 0.018$ & & \\
      $\Sigma^+\to\Lambda^0\ell^+\nu$ & $0.25\pm 0.06$ & & \\ \hline
  \end{tabular}
 \end{center}
 \hspace{10 em} $^*$ Rate in $10^{-3}\,$s$^{-1}$
 \caption[Table 2]{The baryon semi-leptonic decay data used in our analysis.} 
\end{table}

\begin{table}
 \begin{center}
  \begin{tabular}{cc}
   \hline
      $n\to p e\bar\nu$ & $+7.3$ \\
      $\Lambda^0\to p e\bar\nu$ & $+3.9$ \\
      $\Sigma^-\to n e\bar\nu$ & $+1.1$ \\
      $\Xi^-\to\Lambda^0 e\bar\nu$ & $+1.3$ \\
      $\Xi^0\to\Sigma^+ e\bar\nu$ & $+1.3$ \\
      $\Xi^-\to\Sigma^0 e\bar\nu$ & $+1.3$ \\
      $\Sigma^-\to\Lambda^0 e\bar\nu$ & $+1.6$ \\
      $\Sigma^+\to\Lambda^0\bar e\nu$ & $+1.6$ \\ \hline
  \end{tabular}
 \end{center}
 \caption[Table 3]{The radiative corrections to semi-leptonic
 decay rates of baryons in \%.  For $\Xi^0\to\Sigma^+ e\bar\nu$ and
 $\Xi^-\to\Sigma^0 e\bar\nu$ processes, they are made the same value of
 $\Xi^-\to\Lambda^0 e\bar\nu$.} 
\end{table}

\begin{table}
 \begin{center}
  \begin{tabular}{ccccc}
   \hline
      &  & SU(3) symmetry case  & SU(3) breaking case & SU(2) and SU(3) \\
      &  &  &  & breaking case \\ \hline
      $g_1/f_1$ & $n\to p$  & $0.08$ & $0.01$ & $5.3\times 10^{-7}$ \\
      & $\Lambda^0\to p$ & $2.10$ & $0.41$ & $0.18$ \\
      & $\Sigma^-\to n$ & $1.65$ & $0.19$ & $0.38$ \\
      & $\Xi^-\to\Lambda^0$ & $0.62$ & $0.04$ & $0.21$ \\
      & $\Xi^0\to\Sigma^+$ & $0.08$ & $0.21$ & $0.38$ \\ \hline
      Decay & $n\to p e \bar\nu$          & $5.9\times 10^{-3}$ &
              $2.5\times 10^{-3}$ & $2.4\times 10^{-7}$ \\
       Rate  & $\Lambda^0\to p e \bar\nu$ & $1.33$ & $0.07$ & $0.09$ \\
            & $\Lambda^0\to p \mu\bar\nu$ & $0.37$ & $0.51$ & $0.51$ \\
            & $\Sigma^-\to n e \bar\nu$ & $1.01$ & $0.92$ & $0.92$ \\
            & $\Sigma^-\to n \mu\bar\nu$ & $0.03$ & $0.20$ & $0.22$ \\
            & $\Xi^-\to\Lambda^0 e \bar\nu$ & $9.22$ & $1.24$ & $0.55$ \\
            & $\Xi^-\to\Lambda^0 \mu\bar\nu$ & $1.00$ & $0.97$ & $0.91$ \\
            & $\Xi^0\to\Sigma^+ e \bar\nu$ & $0.11$ & $0.14$ & $1.6\times 10^{-3}$ \\
            & $\Xi^-\to\Sigma^0 e \bar\nu$ & $12.70$ & $0.24$ & $4.8\times 10^{-4}$ \\
            & $\Sigma^-\to\Lambda^0 e \bar\nu$ & $0.15$ & $0.84$ & $0.01$\\
            & $\Sigma^+\to\Lambda^0 e^+\nu$ & $0.08$ & $0.12$ & $0.05$ \\ \hline
            & total $\chi^2$ & $30.51$ & $6.09$ & $4.43$ \\ \hline
                     &  $\chi^2/$d.o.f  & $2.77$  & $1.02$  & $0.89$ \\ \hline
  \end{tabular}
 \end{center}
 \caption[Table 4]{The $\chi^2$ values for three cases are listed.} 
\end{table}

\begin{table}
 \begin{center}
  \begin{tabular}{cccc}
   \hline
      parameters    & SU(3) symmetry case  & SU(3) breaking case & SU(3) and SU(2) \\
                    &     &     & breaking case  \\ \hline
      F & $0.475\pm 0.004$ & $0.499\pm 0.001$ & $0.477\pm 0.007$ \\
      D & $0.793\pm 0.005$ & $0.770\pm 0.004$ & $0.835\pm 0.007$ \\
      a & $-$ & $0.454\pm 0.213$ & $0.099\pm 0.049$ \\
      b & $-$ & 0.067$\pm 0.049$ & $0.072\pm 0.005$ \\
      c & $-$ & 0.065$\pm 0.055$ & $0.043\pm 0.005$ \\
      h & $-$ & $-0.099\pm 0.050$ & $-0.031\pm 0.010$ \\
      $\alpha$ & $0$ (putting) & $0$ (putting) & $-0.949\pm 0.200$ \\
      $\beta$  & $0$ (putting) & $-0.205\pm 0.105$ & $-1.301\pm 0.211$ \\
      $V_{ud}$ & $0.976\pm 0.002$ & $0.976\pm 0.002$ & $0.975\pm 0.002$ \\
      $V_{us}$ & $0.222\pm 0.001$ & $0.221\pm 0.002$ & $0.221\pm 0.002$ \\ \hline
  \end{tabular}
 \end{center}
 \caption[Table 5]{The parameter values for all three cases.} 
\end{table}


\vspace{0.8em}

\begin{thebibliography}{99}
\bibitem{EMC}
J. Ashman et al., EM Collab., Phys. Lett. {\bf B206}, 364 (1988);
 Nucl. Phys. {\bf B328}, 1 (1989).
\bibitem{Cabi63}
N. Cabibbo, Phys. Rev. Lett. {\bf 10}, 531 (1963).
\bibitem{Lampe}
For a review on the nucleon spin structure, see
 B. Lampe and E. Reya, Phys. Rep. {\bf 332}, 1 (2000).
\bibitem{HERMES}
A. Airapetian et al., HERMES Collab., Phys. Rev. Lett. {\bf 92}, 012005 (2004).
\bibitem{Ratcliffe04}
For a recent review, see
 P. G. Ratcliffe, Czech. J. Phys. {\bf 54}, B11 (2004).
\bibitem{Ratcliffe97}
P. G. Ratcliffe, in {\it Deep Inelastic Scattering off
 Polarized Targets: Theory Meets Experiment}, edited
 by J. Bl\"umlein et al., (Zeuthen, 1997) hep--ph/9710458.
\bibitem{Ratcliffe99}
P. G. Ratcliffe, Phys. Rev. {\bf D59}, 014038 (1999).
\bibitem{Song}
X. Song, P. K. Kabir, and J. S. McCarthy,
 Phys. Rev. {\bf D54}, 2108 (1996).
\bibitem{Dai}
J. Dai, R. Dashen, E. Jenkins, and A. V. Manohar,
 Phys. Rev. {\bf D53}, 273 (1996).
\bibitem{Mendieta}
R. Flores-Mendieta, E. Jenkins, and A. V. Manohar,
 Phys. Rev. {\bf D58}, 094028 (1998).
\bibitem{Ehrnsperger}
B. Ehrnsperger and A. Sch\"afer, Phys. Lett. {\bf B348}, 619 (1995).
\bibitem{Donoghue}
J. F. Donoghue, B. R. Holstein, and S. W. Klimt,
 Phys. Rev. {\bf D35}, 934 (1987). 
\bibitem{Karl}
G. Karl, Phys. Lett. {\bf B328}, 149 (1994); erratum-ibid {\bf 341}, 449 (1995).
\bibitem{Renton}
P. Renton, {\it Electroweak Interactions --An Introduction to
 the Physics of Quarks \& Leptons}, (Cambridge, 1990).
\bibitem{Weinberg}
S. Weinberg, Phys. Rev. {\bf 112}, 1375 (1958).
\bibitem{Brene}
N. Brene, L. Veje, M. Roos, and C. Cronstr\"om,
 Phys. Rev. {\bf 149}, 1288 (1966).
\bibitem{Sirlin}
A. Sirlin, Nucl. Phys. {\bf B161}, 301 (1979).
\bibitem{Ademollo}
M. Ademollo and R. Gatto, Phys. Rev. Lett. {\bf 13}, 264 (1964).
\bibitem{Garcia85}
A. Garc\'ia and P. Kielanowski, Lect. Notes Phys. {\bf 222}, 1 (1985).
\bibitem{Cabi03}
N. Cabibbo, E. C. Swallow, and R. Winston,
 Ann. Rev. Nucl. Part. Sci. {\bf 53}, 39 (2003).
\bibitem{Hsueh}
S. Y. Hsueh et al., 
 Phys. Rev. {\bf D38}, 2056 (1988).
\bibitem{Garcia92}
A. Garc\'ia, R. Huerta, and P. Kielanowski, Phys. Rev. {\bf D45}, 879 (1992).
\bibitem{Mateu}
V. Mateu and A. Pich, JHEP {\bf 10}, 41 (2005).
\bibitem{PDG}
Particle Data Group, A. B. Balantekin et al., J. of Phys. {\bf G33}, 1 (2006).
\bibitem{KTeV}
A. Alavi-Harati et al., KTeV Collab., Phys. Rev. Lett. {\bf 87}, 132001 (2001).
\bibitem{Bourquin83}
M. Bourquin et al., Z. Phys. {\bf C21}, 27 (1983).
\bibitem{Toth}
K. T\'oth, T. Margaritisz, and K. Szeg\~o,
 Acta Physiol. Acad. Sci. Hung. {\bf 55}, 481 (1984);
T. Margaritisz, K. Szeg\~o, and K. T\'oth, preprint
 KFKI-1982-52.
\bibitem{Bourquin82}
M. Bourquin et al., Z. Phys. {\bf C12}, 307 (1982).
\bibitem{HERMES06}
A. Airapetian et al., HERMES Collab., hep-ex/0609039 (2006).
\bibitem{Sasaki}
S. Sasaki and T. Yamazaki, in {\it 24th International Symposium on Lattice
 Field Theory, Lattice 2006}, (Tucson, 2006), hep--lat/0610082;
 in preparation of a full paper.
\end{thebibliography}
\end{document}